\documentclass[12pt]{article}
\usepackage{amsmath,amsfonts,epsfig}

\advance \textheight by \topskip
\def\mop #1{\mathop{\rm #1}}

\makeatletter
 \@addtoreset{equation}{section}
 \renewcommand\theequation
 {\ifnum\c@section>\z@\thesection.\fi
 \@arabic\c@equation}
\makeatother

\begin{document}

\title{
\begin{flushright}
{\small USACH-FM-01/06}\\[1.0cm]
\end{flushright}
{\bf
Nonlinear Supersymmetry on the Plane in Magnetic
Field and Quasi-Exactly Solvable Systems}}

\author{{\sf Sergey M. Klishevich${}^{a,b}$}%
\thanks{E-mail: sklishev@lauca.usach.cl}
{ \sf and Mikhail S. Plyushchay${}^{a,b}$}%
\thanks{E-mail: mplyushc@lauca.usach.cl}
\\
{\small {\it ${}^a$Departamento de F\'{\i}sica,
Universidad de Santiago de Chile,
Casilla 307, Santiago 2, Chile}}\\
{\small {\it ${}^b$Institute for High Energy Physics,
Protvino, Russia}}}
\date{}

\maketitle

\vskip-1.0cm

\begin{abstract}
The nonlinear $n$-supersymmetry with holomorphic
supercharges is investigated for the 2D system describing
the motion of a charged spin-$1/2$ particle in an external
magnetic field. The universal algebraic structure underlying
the  holomorphic $n$-supersymmetry is found. It is shown
that the essential difference of the 2D realization of the
holomorphic $n$-supersymmetry from the 1D  case
recently analysed by us consists in appearance of the central
charge entering non-trivially into  the superalgebra. The
relation of the 2D holomorphic $n$-supersymmetry to the 
1D quasi-exactly solvable (QES) problems  is demonstrated
by means of the reduction of the systems with hyperbolic or
trigonometric form of the magnetic field. The reduction of
the $n$-supersymmetric system with the polynomial
magnetic field results in the family of the one- dimensional
QES systems with the sextic potential. Unlike the original 
2D holomorphic supersymmetry,  the reduced 1D
supersymmetry associated with $x^6+...$ family is
characterized by the non-holomorphic supercharges of the
special form found by Aoyama et al.
\end{abstract}
\newpage

\section{Introduction}
The nonlinear supersymmetry is one of the new  developments
of  quantum mechanics
\cite{andrian,Nfold,susy-pb,susy-pf,Nfold1,d1,Nfold2,
machol,sfm,ptf,ptch} revealing itself  variously in such
different systems as
the parabosonic \cite{susy-pb} and parafermionic
\cite{susy-pf} oscillator models,
the fermion-monopole system \cite{machol,sfm}, and
the $P,T$-invariant
systems of planar fermions \cite{ptf} and Chern-Simons
fields \cite{ptch}. Being a natural generalization of the
usual supersymmetry \cite{wit,wit1,susyqm,cooper}, it  is
characterized by the polynomial superalgebra resembling the
nonlinear finite $W$-algebras \cite{Walg}.

A simple universal algebraic structure with
oscillator-like bosonic and oscillator fermionic
variables underlies  the  usual ({\it linear})
supersymmetry in classical and quantum mechanics
\cite{wit,wit1,susyqm,cooper}.
The classical nonlinear supersymmetry with
holomorphic supercharges \cite{susy-pb}
has also  the transparent algebraic structure
related to the $n$-fold mapping of the
complex  plane  associated with the
oscillator-like bosonic variables \cite{d1}.
In what follows we will
refer to  the nonlinear supersymmetry
generated by the holomorphic supercharges
with  Poisson bracket (anticommutator)
being  proportional to the  $n$-th order polynomial
in  the Hamiltonian as to the
{\it holomorphic $n$-supersymmetry}.
However,  the attempt to quantize
the nonlinear supersymmetry
immediately faces the problem of the quantum anomaly.
The quantization of  the one-dimensional systems
was investigated  by us in detail in  Ref.~\cite{d1}
where we showed that  the anomaly-free
quantum systems with holomorphic $n$-supersymmetry
turn out to be closely related to the  quasi-exactly
solvable (QES) systems \cite{turbiner,shifman,ushv,turb}.

This  paper  is devoted to generalization
of the analysis of Ref. \cite{d1} for the case
of two-dimensional systems.
It  will allow us not only to find a universal
algebraic structure underlying the holomorphic
$n$-supersymmetry at the quantum level,
but also to demonstrate  a nontrivial relation  of the
holomorphic
$n$-supersymmetry to  the
non-holomorphic nonlinear  supersymmetry
of  Aoyama et al \cite{Nfold2},
and to establish  the relationship of the $2D$
holomorphic $n$-supersymmetry
with the family of QES  systems
with sextic potential \cite{turbiner,shifman,turb,dunne}
not comprised by the  $1D$  holomorphic $n$-supersymmetry.
Nowadays, this  special  class of QES systems attracts
attention  in the context of the $P,T$-invariant
quantum mechanics 
\cite{nelson,pti,pti1,pti2,tateo,pti3,suzuki}.

The paper is organized as follows. In Section \ref{class} we
consider  the holomorphic  $n$-supersymmetry
realized in the classical 2D system
describing the motion of a charged spin-$1/2$ particle
in an external magnetic field.
Section \ref{qbc} illustrates
the simplest anomaly-free quantum realization
of the holomorphic  $n$-supersymmetry
in  the case of the constant magnetic field.
Section \ref{gmf} is devoted to investigation
of the general aspects of the anomaly-free
quantization of the holomorphic $n$-supersymmetry.
We show that the
quantum mechanical $n$-supersymmetry can be realized
only for magnetic field of special configurations of
the exponential and quadratic form.
Here we  find the universal algebraic structure
underlying the holomorphic $n$-supersymmetry.
The nonlinear superalgebra with
the central charge is discussed in Section \ref{salg},
where we consider also  the reduction of the
2D systems with the exponential magnetic field to
the 1D  systems with the  holomorphic
$n$-supersymmetry. In Section \ref{poly} we show that
the spectral problem  of
the 2D system with the quadratic magnetic field
is equivalent to that of the  1D QES systems with the sextic
potential,
and observe the  relation of the 2D holomorphic
$n$-supersymmetry to the non-holomorphic ${\cal N}$-fold
supersymmetry \cite{Nfold2}.
In Section \ref{disc} the brief summary
of the obtained results is presented and some open problems
to be interesting for further investigation
are discussed.

%%%%%%%%%%%%%%%%%%%%%%%%%%%
%%%%%%%%%%
\section{Classical $n$-supersymmetry}
\label{class}

The classical Hamiltonian of a charged
spin-$1/2$
particle ($-e=m=1$)
with gyromagnetic ratio $g$ moving in
a plane and subjected to a
magnetic field $B({\bf x})$
is given by
\begin{equation}\label{Hcl}
 H=\frac 12\boldsymbol{\mathcal P}^2
 + gB({\bf x})\theta^+\theta^-,
\end{equation}
where
$\boldsymbol{\mathcal P}={\bf p}+{\bf A}({\bf x})$,
${\bf A}({\bf x})$ is a 2D gauge potential,
$B({\bf x})=\partial_1A_2-\partial_2A_1$.
The variables $x_i$, $p_i$, $i=1,2,$
and complex Grassman variables $\theta^\pm$,
$(\theta^+)^*=\theta^-$,
are canonically conjugate with respect to the
Poisson brackets,
$\{x_i,p_j\}_{PB}=\delta_{ij}$,
$\{\theta^-,\theta^+\}_{PB}=-i$.
For even values of the gyromagnetic ratio $g=2n$,
$n\in\mathbb N$,
the system (\ref{Hcl}) is endowed with the
nonlinear $n$-supersymmetry. In this case
the Hamiltonian (\ref{Hcl})
takes the form
\begin{eqnarray}\label{Hz}
 &H_n=\frac 12Z^+Z^- + \frac i2n
 \left\{Z^-,Z^+\right\}_{PB}
 \theta^+\theta^-,&\\
\label{Zcl}
& Z^\pm=\mathcal P_2\mp i\mathcal P_1,&
\end{eqnarray}
which admits the existence of the odd integrals of motion
\begin{equation}\label{Qcl}
 Q^\pm_n=2^{-\frac n2}(Z^\mp){}^n\theta^\pm
\end{equation}
generating
the nonlinear $n$-superalgebra \cite{susy-pb}
\begin{equation}\label{qclas}
 \left\{Q^-_n,Q^+_n\right\}_{PB}=-i(H_n)^n, \qquad
 \{Q^\pm_n,H_n\}_{PB}=0.
\end{equation}
This $n$-superalgebra
does not depend on the explicit form of the
even complex conjugate variables $Z^\pm$. Therefore, in
principle, $Z^\pm$ in generators (\ref{Hz}) and
(\ref{Qcl}) can be arbitrary functions of the bosonic
dynamical variables of the system.

The nilpotent quantity $N=\theta^+\theta^-$
is another obvious even integral of motion.
When the gauge potential ${\bf A}({\bf x})$
is a 2D vector, the system (\ref{Hz}) possesses
the additional even integral of motion
$
 L=\varepsilon_{ij}x_ip_j.
$
The integrals $N$ and $L$ generate
the $U(1)$ rotations of the odd,
$\theta^\pm$, and
even, $Z^\pm$, variables, respectively.
Their linear combination
\begin{equation}\label{J}
 J_n=L+nN
\end{equation}
is in involution with the supercharges,
$\{J_n,Q_n^\pm\}_{PB}$=0,
and plays the role of the central charge of the classical
$n$-superalgebra. As we shall see, at the quantum level
the form of the nonlinear $n$-superalgebra (\ref{qclas})
is modified generically by the appearance of the
nontrivial central charge in the anticommutator
of the supercharges.

%%%%%%%%%%%%%%%%%%%%%%%
\section{Quantum $n$-supersymmetry: constant magnetic
field}
\label{qbc}

We start our investigation of the quantum two-dimensional
nonlinear supersymmetry with considering the simplest case
of the
{\it constant} magnetic field.
The quantum $n$-supersymmetric
Hamiltonian for such a system is
\begin{equation}\label{Bc}
 H_n=\frac 14\left\{Z^+,Z^-\right\}+\frac n2B\sigma_3,
\end{equation}
where $Z^\pm$ are the
quantum
analogues of the variables (\ref{Zcl}) with
${\cal P}_i=-i\partial_i-\frac 12 \varepsilon_{ij}x_jB$,
\mbox{$[{\cal P}_1,{\cal P}_2]=-iB$}, corresponding to
the choice of the symmetric gauge.
Here and in what follows we put $\hbar=1$.
The quantum Hamiltonian (\ref{Bc}) is related to the
classical analogue (\ref{Hz}) via the quantization
prescription
$\theta^\pm=\frac{1}{2}(\sigma_1\pm i\sigma_2)$,
$Z^+Z^-\to\frac 12\{Z^+,Z^-\}$,
$\theta^+\theta^-\to\frac 12[\theta^+,\theta^-]=
\frac 12\sigma_3$.

The system (\ref{Bc}) has the
integrals of motion
$\tilde{\cal P}_i={}-i\partial_i
+\frac 12\varepsilon_{ij}x_jB$,
which are in involution with ${\cal P}_i$ and satisfy the
relation $[\tilde{\cal P}_1,\tilde{\cal P}_2]=iB$. In
terms of the creation-annihilation operators
$$
 a^\pm=\frac 1{\sqrt{2|B|}}\bigl({\cal P}_1\pm
 i\varepsilon{\cal P}_2\bigr), \qquad
 b^\pm=\frac 1{\sqrt{2|B|}}\bigl(\tilde{\cal P}_1
 \mp i\varepsilon\tilde{\cal P}_2\bigr),
 $$
$[a^-,a^+]=1$, $[b^-,b^+]=1$,
$\varepsilon=\mop{sign}B$, the Hamiltonian
(\ref{Bc}) is represented in the form
$$
 H_n=|B|\left(a^+a^-+\frac 12 + \frac n2 \varepsilon
 \sigma_3\right).
$$
Since the Hamiltonian does not depend on
$b^\pm$, the energy levels of the system are infinitely
degenerate. The $n$-supersymmetry of the system
(\ref{Bc}) is generated by the supercharges
\begin{equation}\label{QBc}
 Q^\pm_n=|B|^{\frac n2}\left\{
 \begin{array}{ll}
 (a^\mp){}^n\theta^\pm,&\text{for }B>0,\\[1mm]
 (a^\pm){}^n\theta^\pm,&\text{for }B<0,
 \end{array}
 \right.
\end{equation}
\[
 \left\{Q^-_n,Q^+_n\right\}=
 \left(H_n+\tfrac{n-1}2B\right)
 \left(H_n+\tfrac{n-3}2B\right)
 \ldots
 \left(H_n-\tfrac{n-3}2B\right)
 \left(H_n-\tfrac{n-1}2B\right),
 \qquad [Q^\pm_n,H_n]=0.
\]
Therefore, the present 2D system corresponds to
the $n$-supersymmetric 1D oscillator \cite{susy-pb}.
Due to the axial symmetry of the system,
the operator
\begin{equation}\label{BcJ}
 J_n=\frac 1{B}H_n - \varepsilon b^+b^-
\end{equation}
is (up to an inessential additive constant)
the quantum analogue of
the classical integral (\ref{J}).
However, here the quantum central charge $J_n$
of the $n$-superalgebra plays a secondary role
since it is represented in terms of $H_n$ and integrals
$b^\pm$.

As in the case of the one-dimensional theory \cite{d1},
the attempt to generalize the $n$-supersymmetry
of the system (\ref{Bc}) to the case of the magnetic
field of general form faces the problem of quantum anomaly.
In the next section we show that
the generalization is nevertheless possible
for the magnetic field of special form.
Such a phenomenon has an algebraic foundation
and is similar to that taking place in the $1D$ theory.

\section{Quantum $n$-supersymmetry: general magnetic
field}
\label{gmf}

Here we investigate the general case of the
$n$-supersymmetry with {\it holomorphic}
supercharges for the $2D$ charged particle in
an external magnetic field.

A priori, the quantization prescription that respects the
classical $n$-supersymmetry is not known.
We take the quantum Hamiltonian
of the general form
\begin{equation}
\label{Hg}
 H=\frac 12\boldsymbol{\mathcal P}^2
 +V({\bf x})+L({\bf x})N
\end{equation}
with $\mathcal P_i={}-i\partial_i+A_i({\bf x})$,
$i=1,2$, $N=\theta^+\theta^-=\frac 12(\sigma_3+1)$,
and fix the unknown functions from the condition
of existence of the $n$-supersymmetry.
To begin with, we analyse the $n=2$
supersymmetry, and then will generalize the construction
for arbitrary natural $n$. By analogy with the
one-dimensional
$n=2$ supersymmetry, we consider the second order
odd holomorphic operators \cite{d1}
\begin{equation}\label{q2}
 Q_2^\pm=\frac 12((Z^\mp)^2-q)\theta^\pm,
\end{equation}
where $q\in\mathbb C$ and $Z^\pm$ are
the quantum analogues of (\ref{Zcl}).
These odd operators commute with the Hamiltonian
when
$
 L({\bf x})=2B({\bf x}),
 $
 $
 V({\bf x})={}-B({\bf x}),
$
and magnetic field obeys the equations
\begin{equation}\label{eqB}
 (\partial_1\partial_2+2\mop{Im}q)B({\bf x})=0,
 \qquad
 (\partial_1^2-\partial_2^2-4\mop{Re}q)B({\bf x})=0.
\end{equation}
Let us note that
the expression (\ref{q2}) is the most general
form of the second order holomorphic supercharges
since the
term linear in $Z^\pm$  is excluded from them
by the condition $[H_2,Q^\pm_2]=0$.

The potential $V(x)$ has the pure quantum
nature being proportional to $\hbar$.
Since the resulting Hamiltonian has the form
\begin{equation}\label{h2}
 H_2=\frac 12\boldsymbol{\mathcal P}^2
 +B({\bf x})\sigma_3,
\end{equation}
one can treat $V({\bf x})$
as a quantum correction term
providing the same quantization prescription
as in the case of the constant magnetic field.

In the complex variables
\begin{equation}\label{z}
 z=\frac 12\left(x_1+ix_2\right),\qquad
 \bar z=\frac 12\left(x_1-ix_2\right),
\end{equation}
Eqs. (\ref{eqB}) can be rewritten equivalently as
\begin{equation}\label{eqBz}
 \left(\partial^2-\omega^2\right)B(z,\bar z)=0,
\end{equation}
where $B^*=B$, $\omega^2=4q$,
and the notation $\partial=\partial_z$
($\bar\partial=\partial_{\bar z}$) is introduced.
Below we shall see that
the holomorphic nonlinear $n=2$ supersymmetry given by
the supercharges
(\ref{q2}) and Hamiltonian (\ref{h2})
with magnetic field $B$
defined by Eq. (\ref{eqBz})
admits the generalization for the case of
arbitrary $n\in \mathbb N$
with magnetic field $B$ of the same
structure.

The general solution to Eq. (\ref{eqBz}) is
\begin{equation}\label{Bz}
 B(z,\bar z)=w_+e^{\omega z+\bar\omega\bar z}+
 w_-e^{-\left(\omega z+\bar\omega\bar z\right)}+
 we^{\omega z-\bar\omega\bar z}+
 \bar we^{-\left(\omega z-\bar\omega\bar z\right)},
\end{equation}
where $w_\pm\in\mathbb R$, $w\in\mathbb C$,
$\bar w=w^*$.
On the other hand, for $\omega=0$ the
solution to Eq. (\ref{eqBz}) is the polynomial,
\begin{equation}\label{Bp}
 B({\bf x}) =
 c\left((x_1-x_{10})^2+(x_2-x_{20})^2\right)+c_0,
\end{equation}
with $c$, $c_0$, $x_{10}$, $x_{20}$
being some real constants.

Though the latter solution can be obtained
formally from (\ref{Bz})
in the limit $\omega\to 0$ by rescaling appropriately the
parameters $w_\pm$, $w$,
the corresponding limit procedure is singular
and the cases (\ref{Bz}) and (\ref{Bp})
have to be treated separately.

\begin{figure}[ht]
 \begin{center}
 \epsfxsize=8cm
 \epsfbox{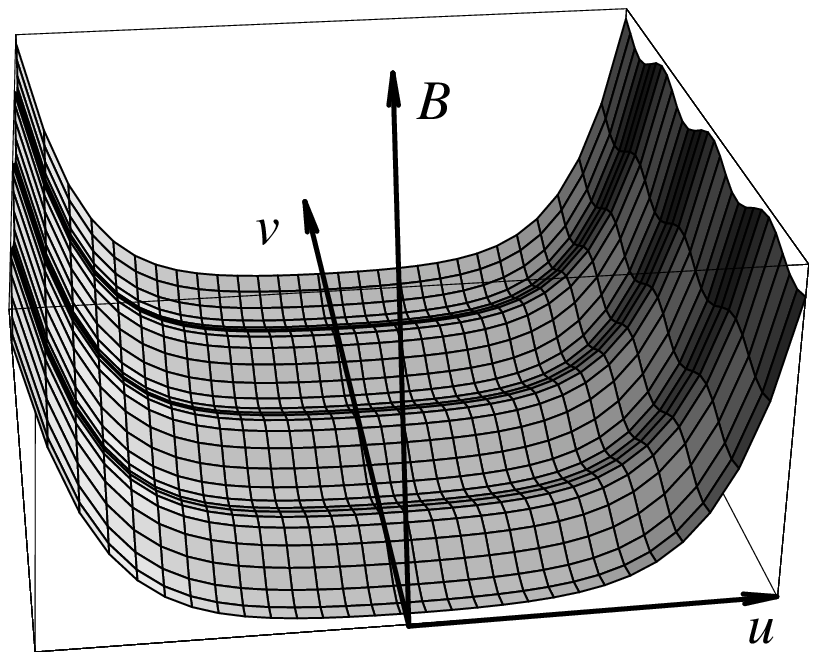}
 \end{center}
 \caption{Magnetic field with
 $\mathop{sign}B_u(-\infty)=\mathop{sign}B_u(+\infty)$.}
 \label{Bpic}
\end{figure}

Rewriting the magnetic field (\ref{Bz}) in terms of
the real variables $x_{1,2}$,
we have
$$
 B({\bf x})=w_+\exp({\bf x}\boldsymbol\omega)
 + w_-\exp(-{\bf x}\boldsymbol\omega)
 +w\exp(i{\bf x}\times\boldsymbol\omega)+
 \bar w\exp(-i{\bf x}\times\boldsymbol\omega),
$$
where
${\bf x}\times\boldsymbol\omega=\varepsilon_{ij}x_i
\omega_j$,
and we have
introduced the constant two-dimensional
vector
$\boldsymbol\omega=(\mop{Re}\omega,-\mop{Im}
\omega)$.
The vector
$\boldsymbol\omega$ defines the preferable coordinate
system,
\begin{align}\label{uv}
 u&=\frac 1{|\omega|}\left(\omega z
 +\bar\omega\bar z\right)=
 \frac{{\bf x}\boldsymbol\omega}{|\omega|},&
 v&={}-\frac i{|\omega|}\left(\omega z
 -\bar\omega\bar z\right)=
 \frac{{\bf x}\times\boldsymbol\omega}{|\omega|},
\end{align}
related to the initial one by the rotation.
In the new coordinates the magnetic field is represented in
the form
\begin{equation}\label{Bvu}
 B(u,v)=B_{u}+B_{v},
\end{equation}
where
\begin{align*}
 B_u&=w_+e^{|\omega|u}+w_-e^{-|\omega|u},&
 B_v&=we^{i|\omega|v}+\bar we^{-i|\omega|v}.
\end{align*}
Thus, the magnetic field is hyperbolic in the $u$ direction
and periodic in the $v$ direction. A typical example of the
magnetic field with
$\mathop{sign}B_u(-\infty)=\mathop{sign}B_u(+\infty)$
is depicted on
Fig.~\ref{Bpic}.

For analysing the nonlinear $n$-supersymmetry for
arbitrary $n\in \mathbb N$,
it is convenient to introduce the complex
oscillator-like operators
\begin{align}\label{Z}
 Z&=\partial+W(z,\bar{z}),&
 \bar{Z}&={}-\bar\partial+\bar{W}(z,\bar{z}),
\end{align}
where the complex superpotential is defined by
$ \mop{Re}W=A_2({\bf x})$,
$\mop{Im}W=A_1({\bf x})$.
The operators $Z$, $\bar Z$ obey the relation
\begin{equation}\label{zzb}
 \left[ Z,\bar Z\right] = 2B(z,\bar z).
\end{equation}
The $n$-supersymmetric Hamiltonian has the form
\begin{equation}\label{Halg}
 H_n=\frac 14\left\{\bar Z,Z\right\}+
 \frac n4\left[Z,\bar Z\right]\sigma_3
\end{equation}
generalizing (\ref{h2}) to the case of arbitrary $n$.
Eq. (\ref{eqBz}) can be rewritten as the
algebraic relation
\begin{equation}\label{int}
 \left[Z,\left[Z,\left[Z,\bar Z\right]\right]\right]=
 \omega^{2}\left[Z,\bar{Z}\right],
\end{equation}
which can be treated as an ``integrability condition''
of the nonlinear holomorphic supersymmetry. Eqs.
(\ref{Halg}) and (\ref{int}) allow us to prove algebraically
by the mathematical induction that the supercharges defined
by the relations
\begin{align}\label{Qalge}
 Q^+_{n+2}&=\frac 12\left(Z^2-\left(\tfrac{n+1}2
 \right)^2
 \omega^2\right)Q^+_n,&
 Q_0^+&=\theta^+,& Q_1^+&=2^{-\frac 12}Z
 \theta^+,
\end{align}
are preserved. This recurrent relation
reproduces correctly the supercharges
$Q_2^\pm$ constructed above.

Since the conservation of the supercharges is proved
algebraically, the operators $Z$, $\bar Z$ can have any
nature (the action of $Z$, $\bar Z$ is supposed to be
associative). For example, they can have a matrix structure.
With this observation the nonlinear supersymmetry can be
applied to the case of matrix Hamiltonians
\cite{turb,mat,mat1,mat2,mat3}.

Thus, the introduction of the operators $Z$, $\bar Z$
allows us to reduce the two-dimensional
nonlinear $n$-supersymmetry
to the pure algebraic construction.

%%%%%%%%%%%%%%%%%%%%%%%%%%%
%%5
\section{Superalgebra for $\omega\ne 0$}
\label{salg}

Unlike the case of linear supersymmetry, the form of
nonlinear superalgebra generated by the operators
$Q^\pm_n$ and $H_n$ defined via relations
(\ref{Halg})-(\ref{Qalge})
depends essentially on the concrete representation of the
operators $Z$, $\bar Z$ satisfying the relation (\ref{int}).
We will use only the representation (\ref{Z}). In this case
the $n$-supersymmetric system (\ref{Halg}) with
$\omega\ne 0$ has the central charge
\begin{equation}\label{Jz}
 J_n=-\frac 14\left(\omega^2\bar Z^2
 +\bar\omega^2Z^2\right)
 +\partial B\bar Z
 +\bar\partial BZ - B^2
 +\frac n2\bar\partial\partial B\sigma_3.
\end{equation}
The anticommutator of the
supercharges contains it for any $n>1$.
For example, the $n=2,3,4$ nonlinear superalgebras are
\begin{align}
 \left\{Q_2^-,Q_2^+\right\}&=H_2^2+\frac 14J_2 +
 \frac{\left|\omega\right|^4}{64},
 \notag\\\notag
 \left\{Q_3^-,Q_3^+\right\}&=
 H_3^3+H_3J_3+\frac 14|\omega|^4H_3 +
 2|\omega|^2\left(|w|^2-w_+w_-\right),
 \notag\\[-3.5mm]\label{ao}\\[-3mm]\notag
 \{Q_4^-,Q_4^+\}&=H_4^4 +\frac 52H_4^2J_4 +
 \frac{41}{2^5}\left|\omega\right|^4H_4^2+
 \frac 9{2^4}J_4^2+12\left|\omega\right|^2
 \left(|w|^2-w_+w_-\right)H_4
 \notag\\\notag
 &+\frac{45}{2^7}\left|\omega\right|^4J_4
 + \frac{3^3}{2^{12}}\left|\omega\right|^8
 -\frac 92\left|\omega\right|^4\left(|w|^2+w_+w_-
 \right).
\end{align}

Let us discuss the eigenvalue problem for the Hamiltonian
(\ref{Halg}) with complex superpotential $W(z,\bar z)$
(see the definition (\ref{Z})). The superpotential
corresponding to the magnetic field (\ref{Bz}) is
\begin{align*}
 W\left(z,\bar{z}\right) &=\frac{1}{\bar{\omega}}
 \left( w_{+}e^{\omega z+\bar{\omega}\bar{z}}-w_{-}e^
 {-
 \left( \omega z+\bar{ \omega}\bar{z}\right) }-we^{
 \omega z-\bar{\omega}\bar{z}}+\bar{w}e^{-\left(
 \omega
 z-\bar{\omega}\bar{z}\right) }\right) \\
 &+f(z)+i\int^{\bar{z}}F(z,\bar{\zeta})\,d\bar\zeta,
\end{align*}
where $F(z,\bar\zeta)^*=F(\zeta,\bar z)$ and
$f(z)$ is a holomorphic function. These arbitrary functions
are associated with the gauge freedom of the system.

In general, the zero modes of the supercharge $Q_n^+$
can be
found. For the sake of simplicity we consider the case
$n=2$ with the following zero modes of $Q_2^+$:
\begin{equation}\label{psi}
 \psi=\left(c_+(\bar z)e^{\frac 12\omega z}+
 c_-(\bar z)e^{-\frac 12\omega z}\right)
 e^{\int^{\bar z}\bar f(\bar\zeta)
 \,d\bar\zeta-\int^zW(\zeta,\bar z)\,d\zeta},
\end{equation}
where $\bar f(\bar z)=f(z)^*$. Now, let us look for the
eigenfunctions of the Hamiltonian associated with the zero
modes. Substitution of (\ref{psi}) into the corresponding
stationary Schr\"odinger equation gives the following
coupled equations for $c_\pm(\bar z)$:
\begin{eqnarray*}
 4\left(w_-e^{-\bar\omega\bar z}+
 \bar we^{\bar\omega\bar z}\right)c_+(\bar z)+
 4Ec_-(\bar z)-\omega c_-{}'(\bar z) &=&0,
\\
 4\left(w_+e^{\bar\omega\bar z}+
 we^{-\bar\omega\bar z }\right)c_-(\bar z)+
 4Ec_+(\bar z)+\omega c_+{}'(\bar z) &=&0.
\end{eqnarray*}
This system of differential equations can be reduced to
the Riccati equation for the function
$y(\bar z)=c_+(\bar z)/c_-(\bar z)$:
\begin{equation}\label{y}
 \omega y'+4\left(e^{-\bar\omega\bar z}w_- +
 e^{\bar\omega\bar z}\bar w\right) y^2+8Ey +
 4\left(e^{\bar\omega\bar z}w_+ +
 e^{-\bar\omega\bar z}w\right) =0.
\end{equation}
Hence, the holomorphic supersymmetry allows ones to
reduce
the two-dimensional spectral problem associated with the
zero modes to the one-dimensional differential equation of
the first order. Unfortunately, we have not succeeded in
finding of the general solution to the equation (\ref{y}).
Therefore, in what follows we consider some special cases
of
the exponential magnetic field.

In the preferable coordinate system (\ref{uv}) the
Hamiltonian takes the form
\begin{equation}\label{Hc}
 H_n=\frac 12\mathcal P_u^2+\frac 12\mathcal P_v^2
 +\frac n2B(u,v)\sigma_3,
\end{equation}
where $\mathcal P_u={}-i\partial_u+A_u(u,v)$,
$\mathcal P_v={}-i\partial_v+A_v(u,v)$. For the magnetic
field (\ref{Bvu}) the gauge potential can be chosen in the
form
\begin{equation}\label{sg}
 A_u(u,v)=\frac 1{|\omega|^2}B_v',\qquad
 A_v(u,v)=\frac 1{|\omega|^2}B_u'.
\end{equation}
Let us consider the system (\ref{Hc}) with the {\it reduced}
magnetic field: $B_v=0$ ($w=0$). Then the gauge
(\ref{sg})
is asymmetric and the central charge takes the form
$$
 J_n=-\frac 14|\omega|^2\hat
 p_v^2+|\omega|^2H_n-4w_+w_-,
$$
where $\hat p_v=-i\partial_v$.
Therefore, in this case instead of
$J_n$, the integral
$\hat p_v$ can be considered
as independent central charge.
It generates translations in the
$v$-direction. Then, e.g., in the case $n=2$
the superalgebra is reduced to the form
$$
 \left\{Q^-_2,Q^+_2\right\}=
 \left(H_2+\tfrac{\left|\omega\right|^2}8\right)^2
 -\frac 1{16}|\omega|^2\hat p_v^2-w_+w_-.
$$
In the gauge (\ref{sg}) the Hamiltonian
can be written as
\begin{equation}\label{hn1}
 H_n={}-\frac 12\partial_u^2
 +\frac 12\left(|\omega|^{-2}B_u'
 -i\partial_v\right)^2+\frac n2B_u\sigma_3.
\end{equation}
The coordinate $v$ is cyclic. Representing the wave
functions in the factorised form
$
 \psi(u,v)=e^{ivp_{v}}\psi(u),
$
we reduce the Hamiltonian (\ref{hn1}) to the
one-dimensional
QES Hamiltonian acting on the functions $\psi(u)$:
\begin{equation}\label{hd1}
 H_n={}-\frac 12\partial_u^2+\frac 12W(u)^2
 +\frac n2W'(u)\sigma_3,
\end{equation}
where
\begin{equation}\label{wd1}
 W(u)=\frac 1{|\omega|}\left(w_+e^{|\omega|u}-
 w_-e^{-|\omega|u}\right)+p_v.
\end{equation}
Hence, the $2D$ $n$-supersymmetric system with the
{\it reduced} magnetic field is equivalent to the $1D$
$n$-supersymmetric system. In particular, using the results
of Ref. \cite{d1} on the QES nature  of the
$n$-supersymmetric system (\ref{hd1}) with the
superpotential (\ref{wd1}),
one can calculate
explicitly $n$ ``Landau levels'' and find the corresponding
wave functions in the described reduced case.
One has also to note that, on the other hand,
for some choice of the parameters of the
superpotential (\ref{wd1}), the well-known exactly solvable
system with the Morse potential can be reproduced
\cite{d1}. Hence, in this case one can find  all the
corresponding eigenstates and eigenvalues.
The term ``Landau levels'' is justified here
by the analogy with the case
of the constant magnetic field in which
the Hamiltonian eigenstates
are bounded only in one of
two directions corresponding to
the continuous variables.

The reduced magnetic field with $B_u=0$ ($w_\pm=0$)
can be considered exactly in the same way.
In this case the
resulting $1D$ $n$-supersymmetric system
is characterized by the
trigonometric superpotential.
If $v\in\mathbb R$,
the corresponding $n$ wave functions
which can be found algebraically are not normalizable.
The normalizability can be achieved by
considering $v\in[0,\frac{2\pi }{|\omega|}k]$,
$k\in\mathbb N$,  i.e. by
identifying the initial configuration space as a
cylinder. However, the detailed
consideration of
such a problem lies out of the scope of the present
paper.

%%%%%%%%%%%%%%%%%%%%%%%%%%%
%\newpage

\section{Polynomial magnetic field and $x^6+...$
family of quasi-exactly solvable potentials}
\label{poly}
Let us turn now to the case of the polynomial magnetic
field.
The operator
\begin{equation}\label{Jp}
 J_n=\frac 1{4c}\left(\partial B(z,\bar z)\bar Z
 +\bar\partial B(z,\bar z)Z-B^2(z,\bar z)
 +\frac n2\bar\partial\partial B(z,\bar z)
 \sigma_3\right)
\end{equation}
is the integral of motion of the system (\ref{Halg})
with the magnetic
field (\ref{Bp}).
It can be obtained from the operator $J_n$
(\ref{Jz}) in the limit $\omega\to 0$
via the same rescaling of the parameters of the exponential
magnetic field which transforms (\ref{Bz})
into (\ref{Bp}).
The essential feature of this integral is
its linearity in derivatives.

The polynomial magnetic field (\ref{Bp}) is
invariant under rotations
about the point $(x_{10},x_{20})$.
Therefore, one can expect that the operator
(\ref{Jp}) should be related to
a generator of the axial symmetry.
To use the benefit of this symmetry,
we pass over to
the polar coordinate system
with the center at the point
$(x_{10},x_{20})$.
Then the magnetic field is radial,
\begin{equation}\label{Br}
 B(r)=cr^2+c_0,
\end{equation}
and the Hamiltonian reads as
\begin{equation}\label{Hpol}
 H_n={}-\frac 12\left({\cal D}_r^2
 +r^{-2}{\cal D}_\varphi^2+r^{-1}{\cal D}_r\right)
 +\frac n2B(r)\sigma_3.
\end{equation}
Here ${\cal D}_r=\partial_r+iA_r(r,\varphi)$,
${\cal D}_\varphi=\partial_\varphi+iA_\varphi(r,\varphi)$,
the magnetic field is given by
\begin{equation}\label{Ba}
 B=r^{-1}\left(\partial_rA_\varphi
 -\partial_\varphi A_r\right),
\end{equation}
and the supercharges have the simple structure
(cf. with Eq. (\ref{QBc})):
\begin{equation}\label{Qpol}
Q_n^+=2^{-\frac n2}Z^n\theta^+,\qquad
 Q_n^-=2^{-\frac n2}\bar Z^n\theta^-.
\end{equation}
As in the case $\omega\ne 0$, the anticommutator of the
supercharges (\ref{Qpol}) is a polynomial of the
$n$-th degree
in $H_n$, $\{Q_n^-,Q_n^+\}=H_n^n+P(H_n,J_n)$,
where
$P(H_n,J_n)$ denotes a polynomial of the $(n-1)$-th
degree.
For example, for $n=2,3,4$ we have
\begin{align}
 \{Q_2^-,Q_2^+\} &= H_2^2 + cJ_2,\notag\\\label{ap}
 \{Q_3^-,Q_3^+\} &= H_3^3 + 4cH_3J_3 - 2c_0c,\\
 \notag
 \{Q_4^-,Q_4^+\} &= H_4^4 + 10cH_4^2J_4
 +9c^2J_4^2 - 12c_0cH_4 - 9c^2.
\end{align}
These expressions can be obtained from (\ref{ao}) via
the limiting procedure discussed above.

For the radial magnetic field it is convenient to use the
asymmetric gauge
\begin{equation}\label{Ap}
 A_\varphi=\frac 14cr^4+\frac 12c_0r^2,
 \qquad A_r=0.
\end{equation}
One could add an additive constant to
$A_\varphi$ since this does not affect the magnetic field
(\ref{Ba}). But such a constant would lead to a singular
gauge potential in the Cartesian coordinates owing to the
singular at the coordinate origin nature of the polar
system. By the same reason the constant cannot be
removed by a gauge transformation. Therefore it has to
vanish.

In the gauge (\ref{Ap}),
the Hamiltonian (\ref{Hpol}) is simplified:
\begin{equation}\label{h1}
 H_n={}-\frac 12\left(\partial_r^2+r^{-1}\partial_r
 -r^{-2}\left(A_\varphi^2(r)-2iA_\varphi(r)
 \partial_\varphi -\partial_\varphi^2\right)\right)
 +\frac n2B(r)\sigma_3.
\end{equation}
The angle variable $\varphi$ is cyclic and
the eigenfunctions of (\ref{h1})
can be represented as
\begin{equation}\label{ps1}
 \Psi(r,\varphi)=
 \begin{pmatrix}
 e^{im'\varphi}\chi_{m'}(r)\\[2mm]e^{im\varphi}\psi_m(r)
 \end{pmatrix},\qquad
 m,\ m'\in\mathbb N.
\end{equation}
In the gauge (\ref{Ap}) the integral
$J_n$ takes the form
\begin{equation*}
 J_n={}-i\partial_\varphi-\frac{c_0^2}{4c}
 +\frac n2\sigma_3.
\end{equation*}
Up to a constant, this integral is equal to (\ref{BcJ}).
Moreover, the system with the constant magnetic field
is recovered in the limit $c\to 0$. In this case
$cJ_n\to -c_0^2/4$ that means that $J_n$ disappears from
(\ref{ap}) recovering the corresponding superalgebra for the
system with the constant field.

The simultaneous eigenstates of the
operators $H_n$ and $J_n$
have the structure
\begin{equation}\label{ps2}
 \Psi_m(r,\varphi)=
 \begin{pmatrix}
 e^{i(m-n)\varphi}\chi_m(r)\\[2mm]e^{im\varphi}\psi_m(r)
 \end{pmatrix}
\end{equation}
and satisfy the equation
\begin{equation}\label{Je}
 J_n\Psi_m(r,\varphi) = \left(m-\frac n2
 - \frac{c_0^2}{4c}\right)\Psi_m(r,\varphi)
\end{equation}
Thus, the integral $J_n$
is associated with the axial symmetry
of the system under consideration and is
(up to an additive constant)
the exact
quantum analogue of (\ref{J}).

Since the angle variable $\varphi$
is cyclic, the 2D Hamiltonian
(\ref{h1}) can be reduced
to the 1D Hamiltonian.
The kinetic term of the
Hamiltonian (\ref{h1}) is Hermitian with respect to the
scalar product with the 
measure $d\mu=rdrd\varphi$. In order to obtain a
one-dimensional system with the usual scalar product
defined
by the measure $d\mu=dr$, one has to perform the
similarity transformation
\begin{align}\label{sim}
 H_n&\to UH_nU^{-1},&
 \Psi&\to U\Psi,&
 U&=\sqrt r.
\end{align}
Since the system obtained after such a transformation is
originated from the two-dimensional system,
one should always
keep in mind that the variable $r$ belongs to the half-line,
$r\in[0,\infty)$.

In what follows we refer to the Hamiltonian acting on the
lower component of the state (\ref{ps2}) as the bosonic
Hamiltonian and to that acting on the upper component as
the fermionic one. They form the
$n$-supersymmetric system.

After transformation (\ref{sim}), the reduced bosonic
one-dimensional Hamiltonian is
\begin{equation}\label{x6}
 {\cal H}_n^{(2)}={}-\frac 12\frac{d^2}{dr^2}
 + \frac{c^2}{32}r^6 + \frac{c_0c}8r^4
 + \frac 18\left(c_0^2-2c(2n - m)\right)r^2
 + \frac{m^2-\frac 14}{2r^2}
 -\frac 12(n-m)c_0.
\end{equation}
This Hamiltonian gives (for $c>0$) the well-known family
of the quasi-exactly solvable systems 
\cite{turbiner,shifman,turb,dunne}.
According to the general theory of $1D$ QES systems,
they
are characterized by the weight $j$ that defines the
corresponding finite-dimensional non-unitary representation
of the algebra $sl(2,\mathbb R)$. In the case (\ref{x6}) the
integer parameter $n$ is related to the corresponding
weight
$j$ as $n=2j+1$.

The superpartner ${\cal H}_n^{(1)}$
can be obtained from
${\cal H}_n^{(2)}$ by the substitution $n\to-n$,
$m \to m-n$. The
supersymmetric pair of
the $2D$ Hamiltonians is related to the
corresponding pair of $1D$ Hamiltonians as
\begin{align}\label{h2to1}
e^{-i(m-n)\varphi}UH_n^{(1)}U^{-1}e^{i(m-n)\varphi}
&=
 {\cal H}_n^{(1)},&
 e^{-im\varphi}UH_n^{(2)}U^{-1}e^{im\varphi}
 &=
 {\cal H}_n^{(2)}.
\end{align}
Here we imply that the operator
$\partial_\varphi$ on l.h.s.
acts according to the rule
$\partial_\varphi e^{ik\varphi}=
e^{ik\varphi}ik$.

The supercharges (\ref{Qpol}) are non-diagonal in
$\varphi$
since in the gauge (\ref{Ap})
the operators $Z$, $\bar Z$
have the form
\begin{align*}
 Z&=e^{-i\varphi}\left(\partial_r + \frac 1r(A_\varphi(r)
 -i\partial_\varphi)\right),&
 \bar Z&=e^{i\varphi}\left({}-\partial_r
 +\frac 1r(A_\varphi(r)-i\partial_\varphi)\right).
\end{align*}
The operator $Z$ decreases the angular momentum of the
state in $1$
while $\bar Z$ increases it. Due to this property the
supercharges (\ref{Qpol}) perform the proper mixing of
the upper
and lower states (\ref{ps2}). The reduction of the
corresponding $2D$ differential operators to
$1D$ looks like
\begin{align}\label{z2to1}
 e^{-i(m-n)\varphi}UZ^nU^{-1}
 e^{im\varphi}&={\cal Z}_n,&
 e^{-im\varphi}U\bar Z^nU^{-1}e^{i(m-n)\varphi}&=
 {\cal Z}_n^\dag,
\end{align}
with ${\cal Z}_n$ given by
$$
 {\cal Z}_n=\left(A - \frac{n-1}{r}\right)
 \left(A - \frac{n-2}{r}\right)\ldots A,
$$
where $A=\frac d{dr}+W(r)$ and the superpotential is
\begin{equation}\label{Wx6}
 W(r)=\frac 14cr^3+\frac 12c_0r+\frac{m-\frac 12}r.
\end{equation}
Using the relations (\ref{h2to1}), (\ref{z2to1}) and
$[Q_n^\pm,H_n]=0$, one can
obtain the
one-dimensional intertwining relations:
\begin{align*}
 {\cal Z}_n{\cal H}_n^{(2)}&=
 {\cal H}_n^{(1)}{\cal Z}_n,&
 {\cal Z}_n^\dag{\cal H}_n^{(1)}&=
 {\cal H}_n^{(2)}{\cal Z}_n^\dag.
\end{align*}
Hence, the one-dimensional odd operators
\begin{align}\label{qx6}
 {\cal Q}_n^+&=2^{-\frac n2}{\cal Z}_n\theta^+,&
 {\cal Q}_n^-&=\left({\cal Q}_n^+
 \right)^\dag
\end{align}
are the true supercharges of the $1D$
$n$-supersymmetric system,
\begin{equation}\label{d1s}
[{\cal Q}_n^\pm,{\cal H}_n]=0, \qquad\text{where}\quad
 {\cal H}_n=
 \begin{pmatrix}
 {\cal H}_n^{(1)}&0\\
 0& {\cal H}_n^{(2)}
 \end{pmatrix}.
\end{equation}
The form of the anticommutator of the supercharges can be
obtained from the corresponding two-dimensional case via the
formal substitution $J_n\to m-\frac n2-\frac{c_0^2}{4c}$.
In the one-dimensional $n$-supersymmetric system
(\ref{x6}), (\ref{qx6}) the integer parameter $m$ can be
formally extended to the whole real line ($m\in\mathbb R$).
A  similar prescription is used when the two-particle
Calogero model is treated as that appearing from the
reduction of the 3D oscillator \cite{cooper}. Note that here
the Calogero model can also be obtained
from (\ref{x6}) in the case  $c=0$ corresponding  to
$B=const$.
{}From the point of view of the  1D quasi-exactly solvable
systems, the Hamiltonian (\ref{x6}) with $c>0$ and
$m<\frac 12$ has $n$ bound states which  can be found
algebraically. But from the viewpoint of the nonlinear
supersymmetry, these $n$ algebraic states are the zero modes
of the odd operator $Q^+_n$. The factorised form of the
supercharges allows us to find the explicit form of the zero
modes:
\begin{equation}\label{psk}
 \tilde{\psi}_m(r)=P_{n-1}(r^2)r^{\frac 12- m}
 \exp\left({}-\frac c{16}r^4-\frac{c_0}4r^2\right),
\end{equation}
where $P_n$ is a polynomial of the $n$-th degree 
non-vanishing at zero. The same
form for the algebraic states is given by the
$sl(2,\mathbb R)$ partial algebraization scheme
\cite{turbiner,shifman}. Substituting the combination
(\ref{psk}) into the corresponding stationary Schr\"odinger
equation, the algebraic system of equations for energy and
coefficients in $P_{n-1}(r^2)$ can be obtained.
On the other hand,
the energies of the $n$ levels are the roots of the
polynomial that the anticommutator
of the supercharges is proportional to.

For $m\ge\frac 12$ the supercharge $Q^+_n$ has no zero
modes, and hence, the $1D$ $n$-supersymmetry is
spontaneously broken.

It is necessary to stress that for
$m^2=\frac 14$ the Hamiltonian (\ref{x6}) has
the non-singular sextic potential and hence, in principle
the system can be treated on the whole line,
$r\in\mathbb R$. The approach based on the
finite-dimensional representations of the algebra
$sl(2,\mathbb R)$ allows ones to find exactly $n$ even
bound
states for $m=\frac 12$ and $n$ odd states
for $m=-\frac 12$. But this approach gives no
explanation why the intermediate states of the opposite
parity are omitted. Having in mind the tight
relationship between the Lee-algebraic approach and the
holomorphic $n$-supersymmetry (see also \cite{d1} for the
details), one can say that the explanation lies in
originating the system with the sextic potential from the
two-dimensional $n$-supersymmetric system. Here it is
necessary to note that even for the Hamiltonian (\ref{x6})
with the non-singular potential (at
$m^2=\frac 14$) the corresponding
supercharges are singular at zero.

It is worth emphasizing that in the initial $2D$
supersymmetric system the supercharges are holomorphic
whereas in the reduced $1D$ system they have a
non-holomorphic form. The reduced $1D$ supersymmetric system
belongs to the so called generalized $\cal N$-fold
supersymmetry of type A \cite{Nfold2}, being a generalization
of the one-dimensional holomorphic supersymmetry~\cite{d1}.

We have obtained an interesting picture for the case of the
polynomial magnetic field. The two-dimensional system with
the nonlinear $n$-supersymmetry described in terms of the
holomorphic supercharges (\ref{Qpol}) corresponds to an
infinite number of the one-dimensional supersymmetric
systems (\ref{d1s}) with the non-holomorphic supercharges:
for every  (integer) $m$ satisfying the relation
$m<1$, the $1D$ $n$-supersymmetry is exact, while for
$m\ge 1$ it is spontaneously broken. On the other hand, the
$2D$
holomorphic $n$-supersymmetry is exact. It is characterized
by the infinite-dimensional subspace of zero modes. However,
there is essential difference between the cases $n=1$ and
$n\ge 2$. For the linear supersymmetry the
basic relation $\{Q_1^-,Q_1^+\}=H_1$ means that all the
zero modes have zero energy, and so, the ground state of
such 2D system is infinitely degenerate in $m$. On the
contrary, for the nonlinear supersymmetry with $n\ge 2$ the
corresponding zero modes are non-degenerate in the energy by
virtue of the nontrivial presence of the central charge
$J_n$ in the superalgebra.

%\newpage
\section*{Discussion and Outlook}
\label{disc}
To conclude, let us summarize and discuss the obtained
results and indicate some problems that deserve further
attention.
\begin{itemize}
\item\it
 The holomorphic $n$-supersymmetry of the 2D system
 describing the motion of a charged spin-$1/2$ particle in
 an external magnetic field provides the gyromagnetic ratio
 $g=2n$ both
 at the classical and quantum levels.
\end{itemize}
This is a natural generalization of the well-known
restriction on the value of gyromagnetic ratio ($g=2$)
related to the linear supersymmetry \cite{cooper}. Here it
is necessary to have in mind that saying about the
spin-$1/2$ particle with gyromagnetic ratio $g=2n$, we
proceed from the structure of the $n$-supersymmetric $2D$
Hamiltonian  given by Eqs. (\ref{Halg}), (\ref{zzb}).
However, the complete ($3D$) spin structure does not appear
in the construction, and hence, the $2D$ system
(\ref{Halg}), (\ref{zzb}) could also be interpreted, e.g.,
as the spin-$n/2$, $g=2$ particle with separated
polarizations $s_z=\pm n/2$. Such alternative interpretation
is in correspondence with the relationship between the
nonlinear supersymmetry and parasupersymmetry discussed in
Ref. \cite{susy-pb}.
\begin{itemize}
\item\it
 The algebraic foundation of the holomorphic
 $n$-supersymmetry is ascertained.
\end{itemize}
The Hamiltonian (\ref{Halg}) and the supercharges
(\ref{Qalge}) of any holomorphic supersymmetric system are
defined in terms of the operators $Z$ and $\bar Z$ only. The
``integrability condition'' (\ref{int}) arises at the
quantum level and guarantees the conservation of the
supercharges in a pure algebraic way. Thus, the formulation
of the nonlinear holomorphic supersymmetry does not depend
on representation of the operators $Z$ and $\bar Z$. In this
sense the holomorphic
$n$-supersymmetry can be treated as a direct
algebraic extension of the usual linear supersymmetry. The
one-dimensional representation of the operators $Z$,
$\bar Z$ was explored in Ref.~\cite{d1} while here we have
investigated the realization of the holomorphic
supersymmetry in the systems on the plane. We have found
that there is an essential difference of the two-dimensional
realization from the one-dimensional case.
\begin{itemize}
\item\it
 In the 2D systems the additional integral of motion $J_n$
 has been found (see expressions (\ref{Jz}) and (\ref{Jp})).
 This integral is a central charge that enters non-trivially
 into the nonlinear superalgebra (see equations (\ref{ao})
 and (\ref{ap})).
\end{itemize}
Technically, the holomorphic $n$-supersymmetry facilitates
finding the form of this additional integral. This is the
important point since, in general, such a problem is rather
laborious.

For the 2D systems the ``integrability condition''
(\ref{int}) is represented as the differential equation
(\ref{eqBz}) for the magnetic field. The general solution
has the exponential form (\ref{Bz}) for $\omega\ne 0$, or
the polynomial form (\ref{Bp}) for $\omega=0$. The latter
solution can be formally obtained from the former in the
limit $\omega\to 0$. In the exponential case the magnetic
field has the orthogonal hyperbolic and trigonometric
directions. The oscillating behaviour of the field in one
direction means that the eigenstates of the Hamiltonian are
not normalizable. This situation is similar to the case of
the constant magnetic field. However, for such exponential
configuration of the magnetic field we have not succeeded in
finding the energy levels of the Hamiltonian associated with
the zero modes of the corresponding supercharge.
\begin{itemize}
\item\it
 The systems with the magnetic field of the pure hyperbolic
 or pure trigonometric form have been reduced to the
 one-dimensional problems with the nonlinear holomorphic
 supersymmetry~\cite{d1}. \end{itemize} In
 Ref.~\cite{cooper} a similar reduction of the 2D system
 with linear supersymmetry was considered in the context of
 application of the shape-invariant potentials to the 2D
 spectral problem. \begin{itemize} \item\it The
 $n$-supersymmetric system with the polynomial magnetic
 field (\ref{Bp}) has been reduced to the well-known family
 of one-dimensional QES systems with the sextic potential.
\end{itemize}
This reduction confirms the intimate relationship between
the nonlinear holomorphic supersymmetry and the QES systems
observed in Ref. \cite{d1}. Moreover, it reveals the
nontrivial relation of the holomorphic $n$-supersymmetry to
the non-holomorphic $\cal N$-fold supersymmetry discussed by
Aoyama et al 
\cite{Nfold2}\footnote{The relation of the non-holomorphic
$\cal N$-fold supersymmetry \cite{Nfold2}
with the superpotential  (\ref{Wx6}) to the family of QES
systems with the sextic
potential was also noted in Ref.~\cite{pti2}.}. 
It is interesting to note that
from the point of view of the reduction, the two-dimensional
holomorphic supersymmetric system contains the infinite set
of one-dimensional systems with the non-holomorphic
supersymmetry in the exact and spontaneously broken phases.
Having in mind the observed relationship between the 2D
holomorphic and 1D non-holomorphic supersymmetries, it would
be reasonable to clarify the following question: {\it Is it
possible to treat the 1D non-holomorphic $\cal N$-fold
supersymmetry of the general form \cite{Nfold2} as a
reduction of some holomorphic $n$-supersymmetry realized in
the 2D Riemannian geometry?}

The underlying algebraic structure of the nonlinear
holomorphic supersymmetry allows ones to apply it for
investigation of the wide class of quantum mechanical
systems including the models described by the matrix
Hamiltonians and the models on a non-commutative space. In
Refs. \cite{turb,mat,mat1,mat2,mat3} the matrix Hamiltonians
were
considered in the context of the QES systems. Therefore it
would be interesting to investigate the possible relation of
the matrix realization of the holomorphic supersymmetry to
such systems. It is worth noting that the matrix extension
of the two-dimensional system (\ref{Hg}) corresponds to the
non-relativistic particle in an external non-Abelian gauge
field. In the case of the models on the non-commutative
space \cite{nocom,nocom1}, the action of quantum
mechanical operators is associative that, in principle, is
enough for realizing the holomorphic $n$-supersymmetry
(\ref{Halg})-(\ref{Qalge}).

At present time, the great attention is attracted by the
so-called $PT$-invariant systems 
\cite{nelson,pti,pti1,pti2,tateo,pti3,suzuki}
described by the non-Hermitian Hamiltonians with a real
spectrum. In Refs.~\cite{pts,pts1}, an extension of the
notion of usual supersymmetry was proposed for such systems.
The QES systems have also found an application to this
subject. Since the discussed holomorphic $n$-supersymmetry
inherits the properties of the supersymmetric and QES
systems, it would be interesting to extend the construction
to the case of the $PT$-invariant systems.

%\newpage
\vskip 0.5cm
{\bf Acknowledgements}
\vskip 5mm

The work was supported by the grants 1010073
and 3000006 from FONDECYT (Chile) and by DYCIT
(USACH).

%%%%%%%%%%%%%%%%%%%%%%%%%%%

\end{document}